\documentclass[11pt]{article}

\usepackage{graphicx}
\usepackage{multicol}
\usepackage{color}
\usepackage{epsfig,latexsym}

\definecolor{verdon}{cmyk}{1,0.5,1,0}
\definecolor{blue}{cmyk}{0.8,0.8,0,0.}
\definecolor{red}{cmyk}{0.2,1,1,0.0}
\newcommand{\cita}[1]{{\color{blue} \cite{#1}}}

\def\lapprox{\mathrel{\mathop  {\hbox{\lower0.5ex\hbox{$\sim$}
\kern-1.1em\lower-0.7ex\hbox{$<$}}}}}
\def\gapprox{\mathrel{\mathop  {\hbox{\lower0.5ex\hbox{$\sim$}
\kern-1.1em\lower-0.7ex\hbox{$>$}}}}}

\begin{document}

\title{\color{verdon} How precisely neutrino emission from supernova remnants\\ can be constrained by
gamma ray observations? }

\author{
F. L. Villante$^{1,2}$,  F. Vissani$^{2}$\\
$^1${\small\em Universit\`a dell'Aquila, Dipartimento di Fisica, L'Aquila, Italy}\\
$^2${\small\em  INFN, Laboratori Nazionali del Gran Sasso, Assergi (AQ), Italy}
}

\date{}

\maketitle

\def\abstractname{\color{red}\bf Abstract}
\begin{abstract}
{\footnotesize  
  We propose a conceptually and computationally simple
  method to evaluate the neutrinos emitted  
  by supernova remnants 
  using the observed $\gamma$-ray spectrum.
  The proposed method does not require 
  any preliminary parametrization of the gamma ray flux;
  the gamma ray data can be used as an input. 
  In this way, we are able to propagate easily 
  the observational errors and to understand 
  how well the neutrino flux and the signal in 
  neutrino telescopes can be constrained by 
  $\gamma$-ray data. 
  We discuss the various 
  possible sources of theoretical and 
  systematical uncertainties ({\em e.g.}, neutrino oscillation 
  parameters, hadronic modeling, {\em etc.}), 
  obtaining an estimate of the accuracy of our calculation. 
  Furthermore, we apply our approach to 
  the supernova remnant RX J1713.7-3946,
  showing that neutrino emission is very-well 
  constrained by the H.E.S.S. $\gamma$-ray data:
  indeed, the accuracy of our prediction is limited by theoretical 
  uncertainties.
  Neutrinos from RX J1713.7-3946 can be detected 
  with an exposure of the order ${\rm km}^2\times\,{\rm year}$, 
  provided that the detection threshold in future 
  neutrino telescopes will be equal to about 1 TeV.}
\end{abstract}


\newpage

\section{\sf\color{verdon} Introduction}

Under-water and under-ice neutrino telescopes are
instruments aiming to  discoveries. They could 
reveal an effective acceleration of cosmic rays (CR)
in galactic sources (such a supernova remnants \cita{ginz}
and micro-quasars \cita{carla}) and/or in extragalactic sources 
(such as AGN and gamma ray bursts \cita{halzen}).  
However, at present it is possible to obtain 
reliable expectations only for few of these sources,
such as the supernova remnants (SNR) 
discussed in this paper. 

The idea that CR could originate in supernovae 
has been put forward already in 1934 but the 
first quantitative formulation of the conjecture that the 
young SNRs refurnish the Milky Way 
of cosmic rays, compensating the energy losses, 
is due to Ginzburg \& Syrovatskii \cita{ginz}.
In fact, the turbulent gas of SNRs is a large reservoir of kinetic 
energy and this environment can support diffusive 
shock waves acceleration~\cita{fermi}. 
The theory of CR acceleration
in SNRs is still in evolution, but the generic 
expectations are stable. The CR flux in SNRs is expected 
to have a power law spectrum with spectral index $\Gamma=2.0-2.4$ 
at low energies with a 
cutoff at an energy $E_{\rm c}$ which depends on 
the details of the acceleration mechanism and on the 
age of the system. 
In specific implementations 
the cutoff energy $E_{\rm c}$ can be as large as several PeV 
and, thus, is consistent with the ``knee" in the 
CR spectrum at $E\sim3\times 10^{15}\, {\rm eV}$, 
which is believed to mark the transition from 
galactic to extra-galactic origin of CR~\cita{berez}.

In recent times,  
great progress has been made in the observation of SNRs.
In particular, the High Energy Stereoscopic System (H.E.S.S.) \cita{hessWEB} 
has determined quite precisely the gamma ray spectra 
of few SNRs showing that they extend 
above 10 TeV.    
In the context of Ginzburg \& 
Syrovatskii hypothesis, it is natural to postulate that the 
observed gammas are produced by the decay of $\pi^0$ (and $\eta$) resulting from the collision 
of accelerated hadrons with the ambient medium. New and crucial observations are being collected
and the hadronic origin seems to be favored for certain SNRs, such as 
Vela Jr \cita{vela}  
and RX J1713.7-3946 \cita{rxj,rxjhadr}.\footnote{
See also \cita{waxman} for a recent analysis 
leading to a different conclusion.}
It is not yet possible, however, to exclude that
(part of) the observed radiation is produced by electromagnetic processes.
The definitive proof that SNRs effectively accelerate CR
could be obtained by the observation of high energy neutrinos in neutrino telescopes
presently in operation, 
in construction  
or in project \cita{nutel}. 

As well known, there is a strict connection between photon and neutrino fluxes
produced by hadronic processes in transparent sources (see,  
{\em e.g.},~\cita{gaisserbook}), 
which results from the fact that the same amount of energy is roughly given to $\pi^0$, $\pi^+$ and $\pi^-$ 
in hadronic collisions. On this basis, one can estimate the neutrino fluxes expected from a 
sources with known $\gamma$-ray spectra, trying to identify 
detectable sources and/or to optimize the detection strategies. 
 The gamma-neutrino connection has been described 
 in various recent papers 
\cita{x1,Costantini:2004ap,x2,Vissani:2006tf,kappes,x4,x5,Vissani:2008zz} 
at a different level of accuracy, relying 
on different assumptions on the primary cosmic ray spectrum 
and/or hadronic interaction model. 

In this particular moment, when high energy gamma ray 
astronomy is flourishing and the neutrino telescopes are finally becoming a reality, 
it is clearly important to have solid and transparent predictions. 
We continue, thus, the work started in 
\cita{Vissani:2006tf,Vissani:2008zz,Villante:2007mh}, 
proposing a method to calculate neutrinos fluxes which 
is, at the same time, simple, accurate and model-independent. 
Our results are in essence a straightforward 
applications of standard techniques \cita{lipari}, 
but we believe that that they will be 
useful since they improve the existing calculations in various respects. 
In particular, we provide simple analytic expressions 
for the neutrino fluxes which have a general validity and can be 
applied directly to gamma ray data since they do not require any parametrization
of the photon spectrum. This allows us
to propagate easily the observational errors in the 
gamma ray flux and, thus, to understand 
how well
the neutrino flux and the signal in neutrino telescopes can be constrained by 
$\gamma$-ray data. 
We also discuss 
the various possible sources of theoretical and systematical uncertainties, 
obtaining an estimate of the accuracy of our method. 
This kind of analysis is relevant in the present situation, 
since the number of events expected in neutrino telescopes from SNRs is quite 
low (see, {\em e.g.}, \cita{Vissani:2006tf}) and even small (downward) revisions of the expected signals 
may be important and/or require different detection strategies. 
It is thus important to understand the relevance of different assumptions in the calculations and 
the origin of (apparently) contrasting results appeared in the literature. 

The plan of the paper is as follow. In Sect.~\ref{method} we review the method presented in 
our previous 
paper~\cita{Vissani:2006tf}. The relations obtained in this 
work -- Eqs.~(\ref{allkernels}) -- are physically 
equivalent to those presented in \cita{Vissani:2006tf}. However, they are more compact and more 
convenient for numerical computations since we have been able to recast them in such a 
way that they require only one numerical integration. 
In Sect.~\ref{oscillations}, we discuss the effect of neutrinos oscillations (see \cita{Strumia:2006db} for a review) 
and we discuss the relevance of uncertainty in neutrino mixing parameters for the predicted neutrino flux.
In Sect.~\ref{finalrelations} we present our main results, {\em i.e.}, 
Eqs.~(\ref{numuflux}) and (\ref{antinumuflux}) 
which relate the (oscillated) muon neutrino and antineutrino fluxes to the $\gamma$-ray flux. As it is explained, 
the only necessary information to predict neutrino fluxes 
are the relative production rates of the various mesons in hadronic processes, which 
are robust predictions of hadronic interaction models. 
In this paper, we adopt the results from Pythia \cita{pythia}, estimated 
by using the parametrization of hadronic cross sections presented in \cita{Koers:2006dd}. A comparison with 
SYBILL~\cita{sybill} and DPMJET-III \cita{dpmjet} is performed (when possible) by using the parametrization
and tabulations of hadronic cross sections presented by \cita{Kelner:2006tc} and \cita{Huang:2006bp}.
In Sect.~\ref{fluxes} and \ref{rates} we propose a procedure to predict neutrino fluxes and event rate in
$\nu$-telescopes directly from $\gamma$-ray observational data, and we apply it to RX J1713.7-3946 which is presently 
the best studied SNR.
In Sect.~\ref{conclusions} we summarize our main results.

\section{\sf\color{verdon} The photon-neutrino relation}
\label{method}

The interactions of cosmic ray protons with a hydrogen ambient cloud 
result in the production of mesons which subsequently decay producing 
gamma ray and neutrinos. Both gamma rays and neutrinos depend linearly on 
the flux of the primary cosmic ray. We, thus, expect that a linear relation
also exists between photon and neutrino fluxes, which can formally 
be expressed as:
$$
\nonumber
\Phi_{\nu}[E_{\nu}]=\int \frac{dE_\gamma}{E_\gamma} \; K_\nu[E_{\nu},E_\gamma]\,\Phi_\gamma[E_{\gamma}].
$$
as will be precised just below in this Section.

 In order to determine the kernels $K_\nu[E_{\nu},E_\gamma]$,
 we only need to know the relative number of pions, kaons and 
$\eta$ produced by cosmic rays at any given energy, as it is 
explained in \cita{Vissani:2006tf} and further discussed in the following. 
Let us indicate with $R_i[E]$ the number of $i-$particles 
produced per unit time and unit energy in the cloud. Here, for simplicity, 
we assume that the momentum distributions of 
the various particles are approximately isotropic, so that the differential flux
produced in a detector at a distance $D$ (if the $i-$particle is stable) is simply given by:\footnote{
If this assumption is removed, one has to replace here and in the following:
$$
\frac{1}{4\pi}R_{i}[E] 
\longrightarrow 
\frac{d R_{i}[E,{\bf n}]}{d\Omega}
$$
where $dR_{i}/{ d\Omega}$ 
is the rate of $i-$particles produced per unit energy and unit solid angle,
${\bf n}$ is the unit vector in 
the direction connecting the SNR to the detector and 
we have taken into account that the produced particles are almost 
collinear.}
\begin{equation}
\Phi_i[E]= \frac{R_i[E]}{4\pi D^2}
\end{equation}
Photons are mainly produced by $\pi^{0}$ and $\eta$, according to:
\begin{equation}
\Phi_\gamma[E_\gamma]=
\frac{2}{4\pi D^2} \int_{E_\gamma}^\infty \frac{dE}{E} \; \left(R_{\pi^0}[E]+b_{\eta\gamma} \,R_\eta[E]\right)
\label{PhiG}
\end{equation}
where $b_{\eta\gamma}=0.394$ is the $\eta\rightarrow\gamma\gamma$ branching ratio and the factor 2 takes into 
account that two photons are produced for each $\pi^0$ and $\eta$. 
Neutrinos are, instead, mainly produced by charged pions and charged kaons. We can formally write 
(neglecting neutrino oscillations):
\begin{equation}
\Phi_\nu[E_\nu]=\frac{1}{4\pi D^2} \sum_i \int_{E_\nu}^\infty \frac{dE}{E} \,R_{i}[E] \,\omega_{i\nu}[E_\nu/E]
\label{PhiNu}
\end{equation}
where $i=\pi^+,\pi^-,K^+,K^-$ and $\nu=\nu_e,\overline{\nu}_e,\nu_\mu,\overline{\nu}_{\mu}$. 
The quantity $w_{i\nu}[x]\,dx$, with $x=E_\nu/E$, represents the spectrum of neutrinos $\nu$
produced in the decay chain the $i-$meson.

If we assume that the ratios between the production rates of the various mesons are approximately constant 
(see next section), the above relations can be combined in order to obtain the neutrino fluxes 
as a function of the photon flux. Rel.~(\ref{PhiG}) can be, in fact, inverted obtaining:
\begin{equation}
R_{\pi^0}[E] = -\frac{4\pi D^2}{1 + b_{\eta\gamma} f_\eta}\,\frac{E}{2}\,\frac{d\Phi_\gamma[E]}{dE}
\end{equation}
where $f_\eta = (R_\eta/R_{\pi^0})$. This expression can then be used in Eq.~(\ref{PhiNu}) to obtain:
\begin{equation}
\Phi_{\nu}[E_\nu]=\int^\infty _{E_\nu} \frac{dE_\gamma}{E_\gamma} \; 
K_\nu \left[E_{\nu}/E_\gamma \right]\,\Phi_\gamma[E_{\gamma}]
\label{eq5}
\end{equation}
where:
\begin{equation}
K_\nu[x]= -\frac{1}{2(1+b_{\eta\gamma} f_\eta)} \sum_i f_i\,\frac{d\omega_{i\nu}[x]}{d\ln x}
\end{equation}
and 
\begin{equation}
f_i=R_i/R_{\pi^0}\;\;\mbox{ with }i=\pi^+,\pi^-,K^+,K^-,\eta \,. 
\label{newdef}
\end{equation}
In explicit terms, we can write:
\begin{eqnarray}
\label{allkernels}
\nonumber
K_{\nu_e}[x] &=& \frac{1}{1+f'_\eta}
\left(f_{\pi^+} \, g_{\pi\nu_e}[x] + f'_{K^+} \,g_{K\nu_e}[x] \right)\\
\nonumber
K_{\overline{\nu}_e}[x] &=& \frac{1}{1+f'_\eta}
\left(f_{\pi^-} \, g_{\pi\nu_e}[x] + f'_{K^-} \,g_{K\nu_e}[x] \right)\\
\nonumber
K_{\nu_\mu}[x] &=& \frac{1}{1+f'_{\eta}}
\left(f_{\pi^+} \, h_{\pi}[x] + f_{\pi^-} \,g_{\pi\nu_\mu}[x] + f'_{K^+} \,h_{K}[x] + f'_{K^-} \,g_{K\nu_\mu}[x] \right)\\
K_{\overline{\nu}_\mu}[x] &=& \frac{1}{1+f'_{\eta}}
\left(f_{\pi^+} \, g_{\pi\nu_\mu}[x] + f_{\pi^-} \,h_{\pi}[x] + f'_{K^+} \,g_{K\nu_\mu}[x] + f'_{K^-} \,h_{K}[x] \right)
\end{eqnarray}
where we have defined
\begin{equation}
f'_\eta=b_{\eta\gamma} f_\eta \;\;\;\; f'_{K^\pm} = b_{K\nu} f_{K^\pm}
\end{equation}
and $b_{K\nu}=0.634$ is the branching ratio 
for $K^{\pm}$ semi-leptonic decay.
%
The functions $h_i[x]$ account for neutrinos produced in $\pi^+\rightarrow\mu^+ + \nu_\mu$
and $K^+\rightarrow\mu^+ + \nu_\mu$ (and charge conjugated processes) and are given by
\begin{equation}
\label{piondecay}
h_{i}[x] = \frac{1}{2}\, \delta[x-(1-r_{i})]
\end{equation}
where $r_i = (m_\mu/m_i)^2$ with $i=\pi,\,K$.
%
The functions $g_{i\nu}[x]$ account, instead, for neutrinos produced by muons. They also
encode the information on the energy distribution and polarization of muons produced by pions and kaons 
decay and can be expressed as:
\begin{equation}
\label{muondecay}
g_{i\nu}[x]=\frac{g'_\nu[x]}{2(1-r_i)}+\frac{r_i\, g'''_\nu[x]}{(1-r_i)^2}-\left(\frac{g''_\nu[x/r_i]}{2(1-r_i)}+\frac{g'''_\nu[x/r_i]}{(1-r_i)^2}\right)\theta[r_i-x]
\end{equation}
where the relevant polynomials are:
\begin{eqnarray}
\nonumber
g'_{\nu_\mu}[x] &=& 2(1-x)^2(1+2x) \\
\nonumber
g''_{\nu_\mu}[x] &=& 4(1-x^3)/3 \\
g'''_{\nu_\mu}[x] &=& (1-x)^2(1+2x)/3
\end{eqnarray}
and
\begin{eqnarray}
\nonumber
g'_{\nu_e}[x] &=& 12(1-x)^2x \\
\nonumber
g''_{\nu_e}[x] &=& 4(1-x)^3 \\
g'''_{\nu_e}[x] &=& -2(1-x)^3
\end{eqnarray}
Some details of the derivations and some checks 
of these formulae, in the limit of 
``isospin invariance'', are given in the appendix A of 
\cita{Vissani:2008zz}.

We note that the same kind of approach can be used to calculate the flux of electrons 
and positrons 
{\em produced by hadronic processes} 
in the cloud, which may be relevant to calculate radio synchrotron emission. By considering that the spectral distribution of electrons
produced in $\mu\rightarrow e+\overline{\nu}_e+\nu_\mu$ is identical to that of of $\nu_\mu$, we immediately obtain:
\begin{equation}
\Phi_{e^\pm}[E]=\int^\infty _{E_\nu} \frac{dE_\gamma}{E_\gamma} \; 
K_{e^\pm} \left[E/E_\gamma \right]\,\Phi_\gamma[E_{\gamma}]
\end{equation}
where
\begin{eqnarray}
K_{e^-}[x] &=& \frac{1}{1+f'_{\eta}}
\left(f_{\pi^-} \,g_{\pi\nu_\mu}[x]   + f'_{K^-} \,g_{K\nu_\mu}[x] \right)\\
K_{e^+}[x] &=& \frac{1}{1+f'_{\eta}}
\left(f_{\pi^+} \, g_{\pi\nu_\mu}[x]  + f'_{K^+} \,g_{K\nu_\mu}[x] \right)
\end{eqnarray}

\section{\sf\color{verdon}The effect of neutrino oscillations}
\label{oscillations}

Neutrino telescopes are sensitive to muon neutrino and muon antineutrino fluxes at earth that differ from the fluxes 
produced in the supernova remnants due to the effect of neutrino oscillations. 
Denoting with $\Phi_{\nu_\ell}^0$ ($\ell=e,\mu,\tau$)
the flux in absence of oscillations, we have 
$\Phi_{\nu_\mu}=P_{\mu\mu} \Phi_{\nu_\mu}^0 + P_{e\mu} \Phi_{\nu_e}^0$,
where due to cosmic distances, the oscillation probabilities are
given by the formula 
$P_{\ell\ell'}=\sum_{i=1,2,3} |U_{\ell i}^2| |U_{\ell' i}^2|$,
appropriate for averaged oscillations \cita{pont}.
This formula is symmetric in the exchange $\ell\leftrightarrow \ell'$
and is valid for neutrinos and antineutrinos. 

In order to implement oscillations in our approach, we replace the kernel in Eq.~(\ref{eq5}) with the the ``oscillated" kernel:  
\begin{equation}
\Phi_{\nu_\mu}[E_\nu]=\int_{E_{\nu}}^\infty\, \frac{dE_\gamma}{E_\gamma} \; K^{\rm osc}_{\nu_\mu}[E_{\nu}/E_\gamma]\,
\Phi_\gamma[E_\gamma]
\end{equation}
which is given by:
\begin{equation}
K^{\rm osc}_{\nu_\mu}[x] = P_{\mu\mu}\, K_{\nu_\mu}[x] + P_{e\mu}\, K_{\nu_e}[x] .
\end{equation}
This can be expressed as the sum of several contributions, each arising from a different mechanism for neutrino production:
\begin{eqnarray}
\nonumber
K^{\rm osc}_{\nu_\mu}[x] &=& A_1 \, h_\pi[x] + A_2\, g_{\pi\nu_\mu}[x] + A_3\, g_{\pi\nu_e}[x] +  \\
                         &&  + \, B_1 \, h_K[x] + B_2\, g_{K\nu_\mu}[x] + B_3\, g_{K\nu_e}[x]
\label{kernel}
\end{eqnarray}
with the coefficients $A_i$ and $B_i$ given by:
\begin{eqnarray}
\nonumber
A_1 =P_{\mu\mu}\frac{f_{\pi^+}}{1+f_\eta'};\;\;\;\;\;
A_2 =P_{\mu\mu}\frac{f_{\pi^-}}{1+f_\eta'};\;\;\;\;\;
A_3 =P_{e\mu}\frac{f_{\pi^+}}{1+f_\eta'};\;\;\;\;\;\\
B_1 =P_{\mu\mu}\frac{f_{K^+}'}{1+f_\eta'};\;\;\;\;\;
B_2 =P_{\mu\mu}\frac{f_{K^-}'}{1+f_\eta'};\;\;\;\;\;
B_3 =P_{e\mu}\frac{f_{K^+}'}{1+f_\eta'}.\;\;\;\;\;
\label{coeff}
\end{eqnarray}
A similar expression holds for the muon antineutrino flux. 
Namely, the kernel $K^{\rm osc}_{\overline{\nu}_\mu}[x]$ is obtained from Eq.~(\ref{kernel})
with the replacements $f_{\pi^+}\rightarrow f_{\pi^-}$ and
$f_{K^+}\rightarrow f_{K^-}$ (and vice versa) in Rels.~(\ref{coeff})

 \begin{figure}[t]
\par
\begin{center}
\includegraphics[width=4.cm,angle=270]{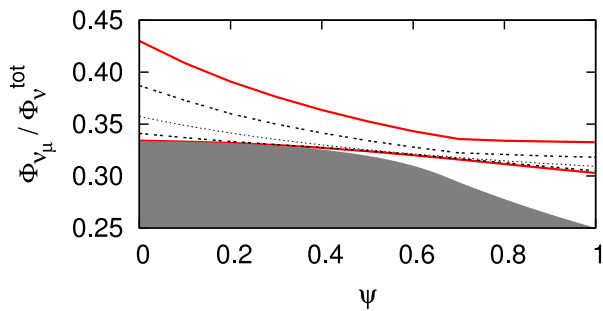}
\end{center}
\par
\caption{\em {\protect\small Suppression of
    cosmic $\nu_\mu$ (or $\bar\nu_\mu$) flux as 
    a function of the electron/muon neutrino (or antineutrino) flux
    ratio. 
    The dashed (continuous) lines enclose 
    the 1$\sigma$ ($2\sigma$) region consistent with the measurements.
    The dotted line identify the most probable value for the
    suppression factor.
    The grey region, instead, is forbidden. }}
\label{prob}
\end{figure}

The oscillation probabilities $P_{\mu\mu}$  
and $P_{e\mu}$ can be evaluated by assuming:
$\theta_{12}=35^\circ \pm 1^\circ$,
$\theta_{23}=42^\circ\pm 4^\circ$ and 
$\theta_{13}=5^\circ \pm 4^\circ$,
close to the range found in a recent 
global analysis of the world data \cita{fogli}.
We obtain the central values:
\begin{equation}
\label{central}
P_{\mu\mu}=0.36 \;\;\;\;\;\; \;\;\;\;\;\; P_{e\mu}=0.26
\end{equation}
that will be used in the following calculations. 
The uncertainties in $P_{\mu\mu}$ and $P_{e\mu}$ are at the level of about  $\sim 10\%$ and 
are almost completely anti-correlated, being mostly due to the spread of $\theta_{23}$ around maximal mixing. 
It is, thus, not immediate to understand how they propagate to the final muon neutrino flux.
Writing:
\begin{equation}
\frac{\Phi_{\nu_\mu}}{\Phi_\nu^{\rm tot}}=
\frac{1}{1+\psi}\left(P_{\mu\mu}+P_{e\mu}\,\psi\right)
\mbox{ where }
\left\{
\begin{array}{l}
\psi=\Phi_{\nu_e}^0 / \Phi_{\nu_\mu}^0 \\[1.5ex]
\Phi_\nu^{\rm tot}=\Phi_{\nu_\ell}^0+\Phi_{\nu_\ell}^0
\end{array}
\right.
\end{equation}
we
see the effect of oscillations 
depends on the neutrino flavor ratio at production $\psi$. 
We also understand that we expect a partial cancellation between the 
contributions of $P_{\mu\mu}$ and $P_{e\mu}$ to the total error budget.

In Fig.~\ref{prob}, we show the l.h.s.~of the above equation as a function of $\psi$. 
The dashed (continuous) lines in the figure 
enclose the 1$\sigma$ ($2\sigma$) region 
consistent with the measurements. 
The dotted line identifies the most probable value for the 
muon neutrino flux suppression factor. 
Several remarks on this figure are in order.

$(i)$~When we consider the 
values $\psi\sim0.5$ characteristic of neutrinos from pion 
decay, we find that the most probable value is $\Phi_{\nu_\mu}/\Phi_{\nu}^{\rm tot}\simeq 0.33$.
This is a well known result, first derived in \cita{pak} and 
often used in the literature 
(see, {\em e.g.}, \cita{kappes,Huang:2007wk}).
For the sake of precision, we note that $\psi$ is the electron/muon neutrino (or antineutrino) flux
ratio {\em at a fixed energy} which may differ from 0.5 even for pion 
decay, due the different energy distribution of the three neutrinos produced 
by pions. If we assume that the neutrino spectra in the SNR is described by a power 
law with spectral index $\alpha=2$ ($\alpha=3$),
we obtain $\psi = 0.54$ ($\psi=0.61$).\footnote{
The neutrino flavor ratio $\psi$ for $K^{\pm}$ decay can largely 
depart from 0.5.
Assuming, again, that the neutrino spectra is described by a power 
law with spectral index $\alpha=2$ ($\alpha=3$),
we obtain $\psi = 0.32$ ($\psi=0.19$). In a more realistic case, in which 
we choose $f_{\pi^+}=f_{\pi^-}=1$ and $f_{K^+}=f_{K^-}=0.1$, we obtain $\psi=0.52$ ($\psi=0.56$).} 
In our approach, however, we do not need to decide {\it a priori} 
the neutrino flavor ratio $\psi$ or the value of $\Phi_{\nu_\mu}/\Phi^{\rm tot}_{\nu}$:
the energy distributions of the produced neutrinos and
the oscillation effects are automatically implemented by Eq.~(\ref{kernel}).

$(ii)$~We note that the possible suppression of the muon neutrino flux is 
bounded from below, as emphasized by the forbidden (grey) region in the figure.
In the extreme case $\psi=0$ we have $\Phi_{\nu_\mu} / \Phi_\nu^{\rm tot}  > 1/3$ 
that can be understood by considering that the (averaged) survival probability  
$P_{\mu\mu}$ cannot be smaller than 1/3; 
if we assume $\psi=1$ we obtain, instead $\Phi_{\nu_\mu} / \Phi_\nu^{\rm tot} = (P_{\mu\mu}+P_{e\mu})/2=(1-P_{\mu\tau})/2>1/4$, 
which follows from the fact that the (averaged) oscillation probability $P_{\mu\tau}$ 
cannot be larger than 1/2. 
The oscillation effect that is realized in Nature happens to be very close to the maximum possible effect, 
namely, to be very close to the forbidden region in Fig.~\ref{prob}. 

$(iii)$~The uncertainty due to the imprecise knowledge of the oscillation 
parameters is at the level of $2\%$ and diminishes with increasing $\psi$,
as a result of the partial cancellation of the much larger (anti-correlated) 
contribution of  $P_{\mu\mu}$ and $P_{e\mu}$ to the total error budget.
It should be noted that uncertainty works 
mostly in the direction to {\em increase} the expected 
muon neutrino (or antineutrino) flux. For  
$\psi\sim 0.5$, we find, in fact, the region
\begin{equation}
\frac{\Phi_{\nu_{\mu}}}{\Phi_{\nu}^{\rm tot}}=(0.33-0.35) \mbox{ at  }2\sigma
\end{equation}
which is extremely asymmetric with respect to the most probable 
value $\sim 0.33$.

\section{\sf\color{verdon} The meson production rates (the $f_i$ factors)}
\label{finalrelations}
 
The last information that we need 
to predict the neutrino fluxes 
are the factors 
$f_i$ 
which give the production 
rates of the various mesons rescaled to that of the neutral pions
(see Eq.~(\ref{newdef})) and can be 
calculated from hadronic interaction models.

The rate of production of $i-$mesons at an energy $E$ is given by:
\begin{equation}
R_{\rm i}[E] = N \int_{E}^\infty 
\frac{dE_{\rm p}}{E_{\rm p}} J_{\rm p}[E_{\rm p}] 
\sigma[E_{\rm p}]
F_{\rm i}\left[\frac{E}{E_{\rm p}},E_{\rm p}\right]
\end{equation}
where $N$ is the total amount of target hydrogen in the cloud, 
$J_{\rm p}[E_{\rm p}]$ is the CR energy spectrum (averaged over the SNR volume, 
see Eq.~(7) in \cita{Villante:2007mh}), and we employed the usual definition of the 
adimensional distribution function $F_i[x,E_{\rm p}]$: 
\begin{equation}
\frac{d\sigma_{i}}{dE} = \frac{\sigma[E_{\rm p}]}{E_{\rm p}} \, 
F_{i}\left[\frac{E}{E_{\rm p}},E_{\rm p}\right] 
\end{equation} 
where $\sigma$ is the total inelastic 
p-p cross section, $E_{\rm p}$ is the proton energy, $d\sigma_i/dE$ is the inclusive cross section for 
$i-$particles production and $E$ represents 
the energy of the produced particle.
Let us assume that the cosmic ray spectrum is roughly described by a power law
in energy. We 
can write $J[E_{\rm p}]\sigma[E_{\rm p}]\propto E_{\rm p}^{-\alpha}$, 
and using the ``quasi-scaling" approximation\footnote{
The ``quasi-scaling" approximation is accurate at the 
few per cent level in the energy range of interest, 
see, {\em e.g.}, \cita{Villante:2007mh}. 
We also remind that the total cross-section $\sigma[E_{\rm p}]$ has a very 
weak dependence on the proton energy, see, {\em e.g.}, \cita{Kelner:2006tc}).}
for hadronic cross sections, {\em i.e.}, taking  
\begin{equation}
F_{\rm i}[x,E_{\rm p}]\rightarrow F_{\rm i}[x]\equiv F_{\rm i}[x,E_{\rm p}^0]
\end{equation}
where $E_{\rm p}^0$ is a fixed reference values for the proton energy,
we finally obtain: 
\begin{equation}
\label{f-factors}
f_i\equiv\frac{R_{i}}{R_{\pi^0}}= \frac{Z_i[\alpha]}{Z_{\pi^0}[\alpha]}
\end{equation}
where:
\begin{equation}
Z_i[\alpha]=\int_{0}^1 dx \; x^{\alpha-1} F_{i}[x]
\end{equation}
The above relation shows that the ratios between the number of produced particles are essentially 
independent on energy and are determined by the $(\alpha-1)-$momenta of the adimensional distribution function $F_i$. 
  

\begin{figure}[t]
\par
\begin{center}
\includegraphics[width=8.0cm,angle=0]{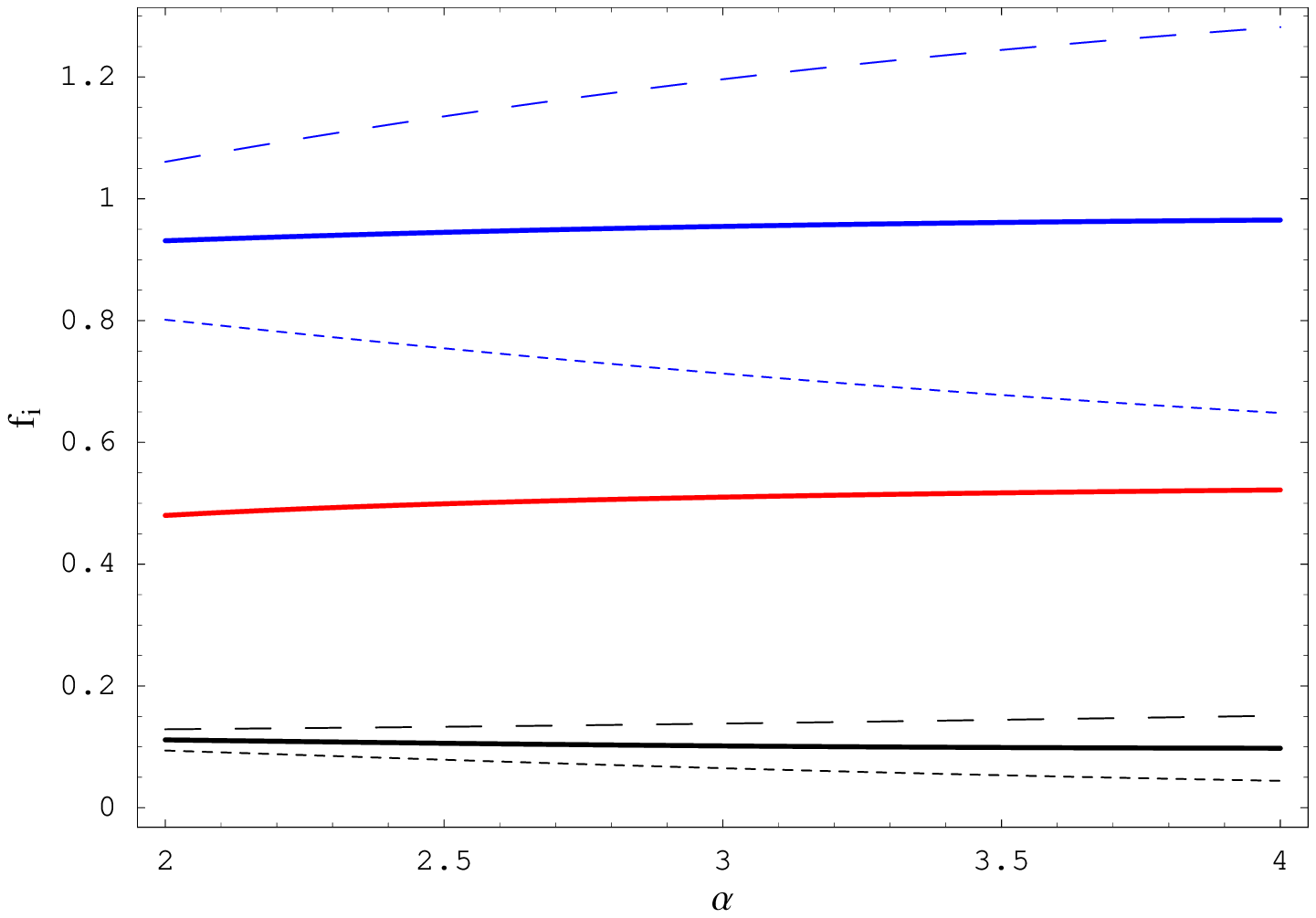}
\includegraphics[width=8.0cm,angle=0]{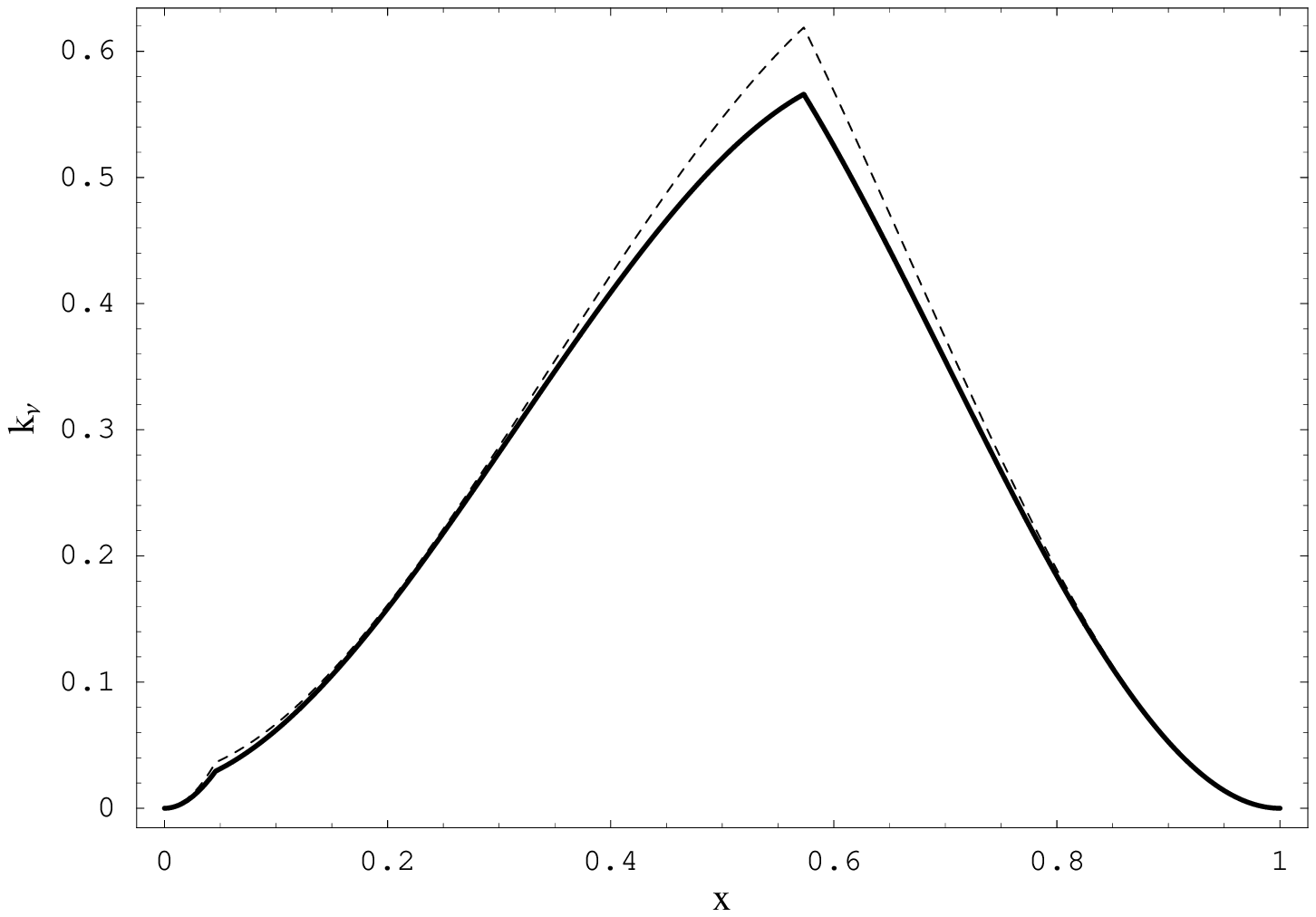}
\end{center}
\par
\vspace{-5mm} \caption{\em {\protect\small {\sc Left panel:} The factors  $f_{\pi^+}$ (dashed blue), $f_{\pi^-}$ (dotted blue), $(f_{\pi^+}\,+\,f_{\pi^-})/2$ (solid blue), $f_\eta$  (red line), $f_{K^+}$ (dashed black), $f_{K^-}$ (dotted black), $(f_{K^+}\,+\,f_{K^-})/2$ (solid black) calculated as a function of $\alpha$ according to Eq.~(\ref{f-factors}). 
{\sc Right Panel:} The kernels $k_{\nu_\mu}[x]$ (solid) and $k_{\overline{\nu}_\mu}[x]$ (dashed).}}
\label{Fig1}
\end{figure}


Presently there are various well developed codes - such as 
Pythia \cita{pythia}, SYBILL \cita{sybill}, QGSJET \cita{qgsjet}, DPMJET-III 
\cita{dpmjet} - which implement different hadronic interaction models 
to study particle production in nucleon-nucleon or nucleon-nucleus collisions.
It would be interesting to make a systematic comparison between the various codes, 
but this is beyond the scope of this work.
In Ref.~\cita{Koers:2006dd}, suitable parametrizations 
of the pion and kaon differential 
cross section in $pp$ collisions from the Pythia code are given.
Moreover, parametrizations of $\pi^0$ and $\eta$ 
productions in $pp$ collisions, according SYBILL and QGSJET code, 
are presented in Ref.~\cita{Kelner:2006tc}. In the left panel of Fig.~\ref{Fig1}, 
we use these parametrizations to calculate the relevant $f_i$ 
factors for $\alpha=2-4$. We see that the quantities $f_i$ are marginally dependent on the assumed 
spectral index and we take the values obtained for $\alpha=2$:
\begin{eqnarray}
\nonumber
f_{\pi^+} = 1.08 \;\;\; & & \;\;\;  f_{\pi^-} = 0.79  \;\;\; \;\;\; \;\;\;  \;\;\;  f_{\eta} =  0.48  \\
f_{K^+}   = 0.13 \;\;\; & & \;\;\;  f_{K^-}   = 0.09 
\label{fi}
\end{eqnarray}
as reference values in our calculations. 
We remark that, using these values, we 
improve on our previous calculation 
\cita{Vissani:2006tf} where we assumed 
``isospin invariance" ({\em i.e.}, $f_{\pi^+}=f_{\pi^-}=1$) and 
we overestimated the rate of $K^{\pm}$ 
production ($f_{K^+}=f_{K^-}=0.2$).

  In order to estimate uncertainties in our method, we should 
compare the $f_{\rm i}$ factors in Eq.~({\ref{fi}}) with 
those obtained by using different hadronic interaction models.
For assumed spectral index $\alpha=2$, the results of \cita{Huang:2006bp}, 
which describes the various possible contribution to $\gamma$-production in 
$p$-ISM collisions, 
can be used to estimate the $f_i$ factors
predicted by DPMJET-III.
In this case, in fact, the factor $Z_i=\langle E_i\rangle/ E_{\rm p}$ coincides with 
the average fraction of the parent particle energy carried by particles of the type $i$.
This quantity can be related to the energy fraction emitted in photons, by
taking into account the branching ratios and the kinematic of the decay processes. 
By using this approach, we estimate (when possible) the values presented in Tab.~\ref{tabella1}, 
where we also show the predictions obtained from Pythia and SYBILL 
parametrizations. 
We remark the Pythia and SYBILL results are obtained by assuming $pp$ collision, 
while the DMPJET-III results refer to the collision of protons with ISM, 
which is assumed to be composed by 90\% protons, 10\% helium nuclei, 0.02\% carbon and 0.04\% oxygen.
 One sees from Tab.~\ref{tabella1} that there is a reasonable agreement between 
the prediction of the different codes. In particular, the $f_i$ 
factors are practically unchanged in the various cases. This means that 
the neutrino-photon ratio 
is a rather solid prediction, which should not suffer from large 
uncertainties in hadronic models and/or in the modelization of the CR flux 
and the ambient medium in the SNR.

Finally,  we discuss the other assumptions in our approach.
Our calculation neglects the contribution of $K^0_S$ and $K^0_L$
(and more rare production channels) to photon and neutrino production. 
According to Tab.~\ref{tabella1}, these processes accounts for several percents increase of 
the photon and neutrino production rates with compensating effects 
in the neutrino-photon ratio, which is affected at the few percents level. 
Moreover, we assumed that the ratio between the production rates of the 
various mesons does not depend on energy, nor on the CR spectral shape. 
This assumption is motivated by the ``quasi-scaling" behavior of hadronic 
cross sections and by the results presented in Fig.~\ref{Fig1} 
and it is valid with
few percents accuracy.
In conclusion, by taking into account 
the approximations implicit in our method 
({\em e.g.}, constant $f_i$ factors, neglected production channels, 
{\em etc.}), 
the uncertainties in hadronic modeling and the uncertainty in neutrino oscillation parameters,
we can safely estimate that the neutrino fluxes predicted by our approach
are accurate at the level of 
about 20\%.\footnote{
As recalled in the Introduction, 
a possible limit of our (and other) calculations is 
the possible presence of a relevant leptonic contamination (component) of the
gamma radiation, that requires efforts in theoretical modeling of the
sources and multi-wavelength observations to be precisely assessed.
Such a possibility implies that, conservatively, we should speak of an 
upper bound on the neutrino signal within the present information.}

\begin{table}
\label{Table1}
\begin{center}
\caption{\em {\protect\small Prediction from different interaction models. See text for details.}}
\vspace{0.5 cm}
\begin{tabular}{l|c|cccccccccc}
										&		&	$\pi^0$	& $\eta$ & $\pi^+$	& $\pi^-$	& $K^+$	& $K^-$	& $K^0_L$	& $K^0_S$		\\
\hline												
\hline
Koers  {\em et al.}~\cita{Koers:2006dd}						&	$Z_i$	&  	0.12	&							& 0.13	& 0.095	& 0.016	& 0.011	& 0.013	&	0.013 \\
{\scriptsize\em {\rm pp} - Pythia}& $f_i$	&  		1		&							& 1.08	& 0.79	& 0.13	& 0.09	& 0.11	&	0.11  \\
\hline
Huang	 {\em et al.}~\cita{Huang:2006bp}								&	$Z_i$	&		0.16	&		0.055$^{\dagger}$			&				&				&	0.019	& 0.014	&	0.016 &	0.017 \\
{\scriptsize\em {\rm p-ISM} - DPMJET-III}	& $f_i$	&  		1		&		0.34$^{\dagger}$			& 			& 			& 0.12	& 0.09	& 0.10	&	0.11   \\
\hline
Kelner {\em et al.}~\cita{Kelner:2006tc}												&	$Z_i$ &	0.13   	&	0.062				&				&				&				&				&				&		\\
{\scriptsize\em {\rm pp}  - SYBILL}	& $f_i$	&  	1						&	0.48				& 			& 			& 			& 			& 			&	  \\
\hline
\end{tabular}\\
\leftline{\footnotesize {}$\;\;\;\;\;\;\;\;\;\;\;\;\;\;\;\;\;\;$$^{\dagger}$Estimated by assuming that "direct $\gamma$ production" in \cita{Huang:2006bp} is due to $\eta$ decays.}
\label{tabella1}
\end{center}
\end{table}

\section{\sf\color{verdon}The neutrino flux}
\label{fluxes}

 By using the $f_i$ factors given in Eq.~(\ref{fi}) and the oscillation probabilities
given in Eq.~(\ref{central}), one can calculate the photon-neutrino kernels 
$K^{\rm osc}_{\nu_\mu}[x]$ and $K^{\rm osc}_{\overline{\nu}_\mu}[x]$
according to Eqs.~(\ref{kernel},\ref{coeff}).
Considering the explicit form of the the various contributions (see Eqs.~(\ref{piondecay},\ref{muondecay})), one obtains the
following simple analytic expression:
\begin{equation}
\Phi_{\nu_\mu}[E] = 0.380 \,\Phi_\gamma[E/(1-r_\pi)]+ 0.0130 \,\Phi_\gamma[E/(1-r_K)]+\int_0^1\frac{dx}{x}\; k_{\nu_\mu}[x]\,\Phi_\gamma[E/x]
\label{numuflux}
\end{equation}
where the first two terms describe neutrinos produced in pions (first term) and kaons (second) decays. 
The kernel $k_{\nu_\mu}[x]$, which takes into account neutrinos produced by muon decay, is shown in the
right panel of Fig.~\ref{Fig1} and it is given by:
\begin{eqnarray}
\nonumber 
k_{\nu_\mu}[x] &=& x^2 (15.34 - 28.93 \, x) \;\;\;\;\;\;\;\;\;\;\;\;\;\;\;\;\;\;\;\;\;\;\;\;\;\;\;\;\;\;\;\;\;\;\;\;\; 
x \le r_K = 0.0458 \\
\nonumber 
               &=& 0.0165 + 0.1193 x + 3.747 x^2 - 3.981 x^3 
               \;\;\;\;\;\;\;\;    
r_K < x < r_\pi \\
\label{knumu}
               &=& (1-x)^2 \,(-0.6698 +  6.588 x ) \;\;\;\;\;\;\;\;\;\;\;\;\;\;\;\;\;\;\;\;\;\;\;\; 
x \ge r_\pi = 0.573               
\end{eqnarray}
A similar relation holds for muon antineutrino flux:
\begin{equation}
\Phi_{\overline{\nu}_\mu}[E] = 0.278 \,\Phi_\gamma[E/(1-r_\pi)]+ 0.0090\,\Phi_\gamma[E/(1-r_K)]+\int_0^1\frac{dx}{x}\; k_{\overline{\nu}_\mu}[x]\,\Phi_\gamma[E/x]
\label{antinumuflux}
\end{equation}
where
\begin{eqnarray}
\nonumber 
k_{\overline{\nu}_\mu}[x] &=& x^2 (18.48 - 25.33 \, x) \;\;\;\;\;\;\;\;\;\;\;\;\;\;\;\;\;\;\;\;\;\;\;\;\;\;\;\;\;\;\;\;\;\;\;\;\; 
x \le r_K = 0.0458 \\
\nonumber 
               &=& 0.0251 + 0.0826 x + 3.697  x^2 - 3.548  x^3 \;\;\;\;\;\;\;\;    
r_K < x < r_\pi \\
\label{kantinumu}
               &=& (1-x)^2 \,(0.0351 +  5.864 x ) \;\;\;\;\;\;\;\;\;\;\;\;\;\;\;\;\;\;\;\;\;\;\;\;\;\; 
x \ge r_\pi = 0.573 
\end{eqnarray}
Eqs.~(\ref{numuflux}) and (\ref{antinumuflux}) 
have a general validity and do not require 
any specific parametrization of the photon spectrum. They will be used in the following to derive the neutrino flux
from the SNR RXJ1713.7-3946 directly from the observational $\gamma-$ray data. 
It is, anyhow, interesting to discuss the application of these relations for the functional forms
most commonly used to parametrize the photon spectrum.

\subsection{\sf\color{verdon}Using parametrized fluxes}

\begin{figure}[t]
\par
\begin{center}
\includegraphics[width=8.5cm,angle=0]{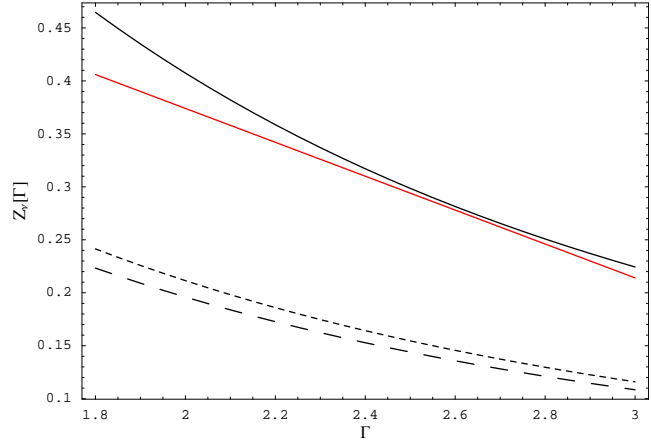}
\end{center}
\par
\vspace{-5mm} \caption{\em {\protect\small The function $Z_\nu[\Gamma]$ for $\nu_\mu$ (dotted), 
$\overline{\nu}_\mu$ (dashed) and $\nu_\mu+\overline{\nu}_\mu$ (solid). The red line is 
the approximate expression $Z_{\nu_\mu+\overline{\nu}_\mu}[\Gamma] = 0.71-0.16\,(\Gamma+0.1)$ 
obtained by Eqs.~(5,6) of \cita{kappes}.}}
\label{suppression}
\end{figure}

If the photon spectrum is approximated with a power law $\Phi_\gamma\propto E^{-\Gamma}$, one immediately sees that:
\begin{equation}
\Phi_\nu[E] = Z_\nu[\Gamma]\cdot\Phi_\gamma[E]
\end{equation}
where:
\begin{eqnarray}
\nonumber
Z_{\nu_{\mu}}[\Gamma]&=&0.380\;x_\pi^\Gamma+0.0130\;x_K^\Gamma+\int_0^1dx \,k_{\nu_\mu}[x]\,x^{\Gamma-1}\\
Z_{\overline{\nu}_{\mu}}[\Gamma]&=&0.278\;x_\pi^\Gamma+0.0090\;x_K^\Gamma+\int_0^1dx \,k_{\overline{\nu}_\mu}[x]\,x^{\Gamma-1}
\end{eqnarray}
with $x_\pi=1-r_\pi=0.427$ and $x_K=1-r_K = 0.954$.
The functions $Z_\nu[\Gamma]$ are the $(\Gamma-1)$-momenta of the photon-neutrino kernels and can be calculated analytically by using Eqs.~(\ref{knumu}) and (\ref{kantinumu}). They are shown in Fig.~\ref{suppression} for 
where the dotted, dashed and solid lines corresponds to $\nu=\nu_{\mu}, \; \overline{\nu}_\mu\;{\rm and}\; (\nu_{\mu} \,+\, \overline{\nu}_\mu) $, respectively. 
The red line is the approximate expression $Z_{\nu_\mu+\overline{\nu}_\mu}[\Gamma] = 0.71-0.16\,(\Gamma+0.1)$ 
obtained by Eqs.~(5,6) of \cita{kappes} where the relationship between $\gamma$-ray and neutrino 
spectra produced by simply parametrized primary proton spectra was studied. 
One sees that our calculation predicts $\sim 10\%$ larger neutrino fluxes. This is not surprising considering that 
the calculation of \cita{kappes} (which is based on
\cita{Kelner:2006tc}) do
not include kaons (and assume equal production 
of $\pi^+$, $\pi^-$ and $\pi^0$).\footnote{
We do not compare 
with the results of the recent publication \cita{Huang:2007wk} because 
they do not provide sufficient details 
to reproduce their results. In particular, 
we are unable to reproduce the line labeled ``Vissani, 2006" in their
Fig.~2 
which, in the intention 
of the authors of \cita{Huang:2007wk}, should amount to an application
of \cita{Vissani:2006tf}.}

The photon spectrum is expected to have a cutoff at high energies. Thus, we can write  
$\Phi_\gamma = N \,\xi_\gamma[E]\,E^{-\Gamma}$, 
where $N$ is a normalization factor and the adimensional function $\xi_\gamma[E]$ 
which modulates the photon spectrum is normalized to 1 at low energies. 
If $\gamma$-rays originate from hadronic processes, the cutoff cannot be too sharp,
since the spectral features of the parent CR spectrum are diluted in
hadronic cascades, as was pointed out in \cita{Villante:2007mh}. If the function 
$\xi_\gamma[E]$ is sufficiently smooth, 
we can extract it from the integrals in Eqs.~(\ref{numuflux}) and (\ref{antinumuflux}), obtaining:
\begin{equation}
\Phi_\nu[E] = N\, Z_\nu[\Gamma] \, \xi_\nu[E]\,E^{-\Gamma}
\end{equation}
with
\begin{equation}
\label{approximatenu}
\begin{array}{c}
\displaystyle
\xi_{\nu_{\mu}}[E] \simeq \frac{1}{Z_{\nu_\mu}[\Gamma]}\left( 0.380\,{x_\pi}^\Gamma\,\xi_\gamma\left[E/x_\pi\right]+
0.013\;{x_K}^\Gamma \, \xi_\gamma\left[E/x_K\right]
+ \xi_\gamma\left[E/x_{\rm m} \right] \int_0^1 dx
\,k_{\nu_\mu}[x]\,x^{\Gamma-1}\right) \\[2ex]
\displaystyle
\xi_{\overline{\nu}_{\mu}}[E] \simeq 
\frac{1}{Z_{\overline{\nu}_{\mu}}[\Gamma]}
\left( 0.278\,{x_\pi}^\Gamma\,\xi_\gamma\left[E/x_\pi\right]+
0.009\;{x_K}^\Gamma \, \xi_\gamma\left[E/x_K\right]
+ \xi_\gamma\left[E/x_{\rm m} \right] \int_0^1 dx \,k_{\overline{\nu}_\mu}[x]\,x^{\Gamma-1}\right)
\end{array}
\end{equation}
where $x_{\rm m}=0.59$ 
approximately coincides with  the maximum of the function 
$k_{\nu_\mu}[x]\,x^{\Gamma-1}$ when $\Gamma \simeq 2-3$. The above relations shows that, 
beside being suppressed by a factor $Z_\nu [\Gamma]$ with respect to the photon flux, 
the neutrino fluxes are also shifted to lower energies. For $\Gamma=2$ the numerical factors of the
various terms in the r.h.s.~of 
Eqs.~(\ref{approximatenu})  become 0.069, 0.012, 0.130
and 0.051, 0.008, 0.137 respectively, showing that the dominant contribution
is provided by neutrino produced in muon decays ({\em i.e.}, last terms in
the r.h.s.~of Eqs.~(\ref{approximatenu})). 
As a consequence, the features in the photon spectrum, such as the presence of a cutoff at an energy $E_{\gamma,\rm c}$, 
are essentially reproduced in the neutrino spectrum at an energy lower by a factor 
$E_{\nu,\rm c}/E_{\gamma,\rm c} = x_{\rm m}\simeq 0.59$, as also noticed by \cita{kappes}.


\subsection{\sf\color{verdon}Using raw data}

RXJ1713.7-3946 is presently the best studied SNR. It has been observed by H.E.S.S. 
during three years from 2003 to 2005 \cita{rxj}.
The data expands over three decades, 
exploring the energy interval $E_\gamma = 0.3-300$ TeV.
The energy resolution of the experiment is equal to
about 20\% and the photon spectrum is sampled in 25 bins 
$\delta E_\gamma/E_\gamma \simeq 0.2$ plus three larger bins at 
high energy. In a previous paper \cita{Villante:2007mh}, we have discussed the information that the 
observational data provide on the parent CR spectrum in the SNR.
Here, we use the observational data to calculate the $\nu_\mu$ and $\overline{\nu}_\mu$
fluxes emitted by this object and to estimate the event rate expected 
in neutrino telescopes.
The proposed procedure, which does 
not require any parametrization of the photon flux, 
is general and can be applied to any other $\gamma$-transparent source.

 As a first step, we ``rescale" the photon and neutrino fluxes according to:
\begin{eqnarray}
\varphi_\gamma[E] &\equiv&  \Phi_\gamma[E] \cdot E^\alpha \\
\varphi_\nu[E] &\equiv&  \Phi_\nu[E] \cdot E^\alpha
\end{eqnarray}
where $\nu=\nu_\mu,\,\overline{\nu}_\mu$. For a proper choice of the parameter $\alpha$, 
the ``rescaled" fluxes are expected to vary slowly with energy. In the following, we adopt:
\begin{equation}
\alpha = 2.5
\end{equation}
which is particularly appropriate for RX J1713.7-3946 (see, {\em e.g.},
Fig.~3 
of \cita{Villante:2007mh} and related discussion).

 We indicate with $\varphi_j \pm \Delta \varphi_j$ the (rescaled) photon flux {\it measured}
in the $j-$th energy bin, centered at a photon energy $E_j$ and covering the energy range 
$(E_{j,{\rm inf}},E_{j,{\rm sup}})$. 
We can approximate the photon flux by:
\begin{equation}
\varphi_\gamma[E_\gamma]=\sum_j \varphi_j \, W_j[E_\gamma]
\label{rawdata}
\end{equation}
where $W_j[E_\gamma]$ are rectangular functions which describes the various energy bins ({\em i.e.}, 
$W_j[E_\gamma]\equiv1$ for $E_{j,\rm inf}\le E_\gamma \le E_{j,\rm sup}$ 
and zero elsewhere).
We immediately obtain 
from Eqs.~(\ref{numuflux},\ref{antinumuflux}) the relation:
\begin{eqnarray}
\nonumber
\varphi_{\nu_\mu}[E] &=& \sum_j \varphi_j \, w_j[E] \\
\varphi_{\overline{\nu}_\mu}[E] &=& \sum_j \varphi_j \, \overline{w}_j[E]
\label{nuflux-rawdata}
\end{eqnarray}
where:
\begin{eqnarray}
\nonumber
w_{j}[E]&=&0.0453\,W_j[E/(1-r_\pi)]+0.0116\,W_j[E/(1-r_K)]+\int_{0}^{1}\frac{dx}{x}\; 
k_{\nu_\mu}[x]\; x^\alpha \, W_j[E/x]\\
\overline{w}_{j}[E]&=&0.0331\,W_j[E/(1-r_\pi)]+0.0080\,W_j[E/(1-r_K)]+\int_{0}^{1}\frac{dx}{x}\; 
k_{\overline{\nu}_\mu}[x]\; x^\alpha \, W_j[E/x]
\label{rawkernel}
\end{eqnarray}

 Rels.~(\ref{nuflux-rawdata}) give the neutrino fluxes as a linear combination 
of the observational values $\varphi_{j}$ of the photon flux.
The functions $w_j[E]$ and $\overline{w}_j[E]$ 
describe the contribution that each 
data point gives to the reconstructed neutrino flux at the energy $E$.
The uncertainty in the neutrino fluxes can be evaluated by propagating 
linearly the observational errors $\Delta \varphi_j$. We obtain: 
\begin{equation}
\frac{\Delta \varphi_{\nu_\mu}[E]}
{\varphi_{\nu_\mu}[E]} = 
\frac{\sqrt{\sum_j  \Delta\varphi_j^2  \,w_j[E]^2}}
{\sum_j \varphi_j  \,w_j[E]}
\label{error}
\end{equation} 
Similarly, the correlation between the values of the $\nu_\mu$ flux 
at two different energies can be calculated by:
\begin{equation}
\varrho_{\nu_\mu}[E,E'] = 
\frac{\sum_k  \Delta\varphi_k^2  \ w_k[E]\ w_k[E']}
{\sqrt{\sum_j  \Delta\varphi_j^2  \ w_j[E]^2}\
 \sqrt{\sum_l  \Delta\varphi_l^2  \ w_l[E]^2}} 
\label{correlation}
\end{equation} 
Analogous expressions can be obtained for $\overline{\nu}_\mu$ flux by replacing
$w_j[E]\rightarrow \overline{w}_j[E]$ in the above expressions.


\begin{figure}[t]
\par
\begin{center}
\includegraphics[width=8.5cm,angle=0]{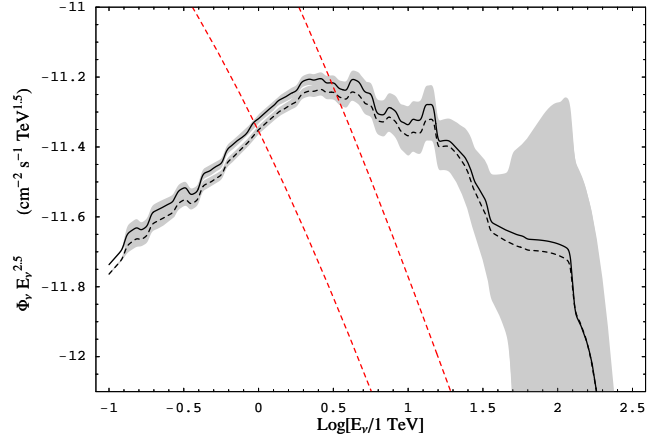}
\end{center}
\par
\vspace{-5mm} \caption{\em {\protect\small The $\nu_\mu$ (solid) and $\overline{\nu}_\mu$ (dotted) flux expected from the SNR RX J1713.7-3946 
according to the H.E.S.S. $\gamma-$ray data. The shaded area are obtained by propagating the observational uncertainties 
in the $\gamma$-ray data. The red dotted lines 
correspond to the atmospheric neutrino 
flux in the vertical direction integrated 
over an angular window $0.5^\circ$ (lower line) 
and $1.0^\circ$ (upper line).\label{fignufluxes}}}
\end{figure}


 The results of the proposed method are displayed in Fig.~\ref{fignufluxes} where we show with solid (dashed) line 
the $\nu_\mu$ flux ($\overline{\nu}_\mu$ flux) expected from RX J1713.7-3946 according to the 
H.E.S.S. observational $\gamma-$ray data. The H.E.S.S. data cover the energy range 
$E_\gamma = 0.3-300$ TeV. In this interval we have described the photon flux 
according to Eq.~(\ref{rawdata}) and we have not relied on theoretical 
assumptions. 
It is, however, unnatural to assume that the photon spectrum vanishes 
outside the probed region. 
For this reason, we have continued 
the photon spectrum at low energy ($E_\gamma\le 0.3$ TeV) assuming 
$\Phi_\gamma = I (E/{\rm 1 TeV})^{-\beta}$ where 
$\beta=2$ and $I=2.47 \cdot 10^{-11}/({\rm cm}^2 \; {\rm s}\;{\rm TeV})$.  
The assumed low-energy photon spectrum smoothly connects with the 
low energy bins observed by H.E.S.S experiment (see, 
{\em e.g.}, \cita{rxj,Villante:2007mh}). 
In order to make minimal assumptions we have not extrapolated 
the photon spectrum  at high energies. The relevance of the assumed 
low and high energy behavior can be studied by considering different 
extrapolations for the gamma ray flux outside the region probed 
by H.E.S.S. experiment. In this way, we have been able to verify the spectral region 
directly constrained by the data is $E_\nu \simeq 0.3 - 100\ {\rm TeV}$.

We can see that the neutrino and antineutrino spectra are well described by 
a power law with spectral index $\gamma\simeq 2$ at low energies and a 
cutoff/transition region at $E_\nu\sim 3-5$ TeV. The ratio $\Phi_{\overline{\nu}_\mu}/\Phi_{\nu_\mu}$ 
is nearly constant and equal to about 0.93. 
The shaded areas describe the {\em observational} errors in the neutrino fluxes and 
are obtained by propagating the uncertainties in the gamma ray data. The neutrino fluxes are well
constrained at low energies where the relative uncertainty is at few per cents level. 
The information degrades at high energy. We have $\Delta \Phi_\nu/\Phi_\nu\sim 30\%$ at 
$E_\nu = 30$ TeV and much worse at larger energies. We remind, for completeness, that 
one has to consider also a {\em systematic} error in our calculation equal to about $20\%$
due to uncertainties in hadronic cross section, neutrino 
oscillation parameters and the approximations implicit in our method,
as conservatively estimated in the previous section.\footnote{
The wiggly behavior of the neutrino spectra is not physically significant. 
It reflects the statistical fluctuations of the photon data in the simple 
interpolation scheme proposed in Eq.~(\ref{rawdata}).
To reduce the effect, we have applied a Gaussian smearing to the neutrino fluxes
predicted by Eq.~(\ref{nuflux-rawdata}) on scales smaller than 
$\delta E_{\nu}/E_{\nu}\simeq0.03$. This is equivalent to smooth the photon flux
on scales $\delta E_{\gamma}/E_{\gamma}\le 0.03$ and 
does not erase any significant features. The smoothing scale is, in fact,
 much smaller than the H.E.S.S. energy resolution (equal to about $20\%$)
 and than the size of the energy bins.}

 The red dotted lines in Fig.~\ref{fignufluxes} show the atmospheric neutrino flux 
 which provides a diffuse background for SNR neutrino detection. 
 The relevance of this background is, clearly, reduced if one is able to 
 observe the source with a good pointing accuracy. The lower (upper) 
 red dotted line is obtained by integrating the muon neutrino flux in 
 the vertical direction given by \cita{lipariatmo} over an a 
 angular window $0.5^\circ$ ($1.0^\circ$), which correspond to the angular resolution
 of neutrino telescopes at energies equal to about 1 TeV. We remind that,
 below $E_{\nu} \sim 1 {\rm TeV}$, the angular response of the detector
 is mainly determined by kinematic of the detection process and, thus, cannot be improved.
 We see that the spectral region in which the signal is expected 
to be larger than the background is above $\sim 1 {\rm TeV}$ quite close to cutoff 
in the $\nu$ spectrum. The red lines in Fig.~\ref{fignufluxes} should 
 be intended as lower limits of the atmospheric $\nu$ background. The atmospheric 
 neutrino flux, in fact, depends on the zenith angle and increase by 
 about one order of magnitude in the horizontal direction. Moreover,
 the angular dimension of the galactic SNRs can be comparable 
 with the detector angular resolution. The SNR RX J1713.7-3946, as an example,
 subtends an angle in the sky equal to about $1.0^\circ$.
   
In the next section, we will calculate the event rate produced by RX J1713.7-3946 
in a telescope located in the Mediterranean sea. We conclude this section by discussing 
whether we can expect neutrino fluxes much larger than that from RX J1713.7-3946. 
In this respect, it was noted in \cita{Villante:2007mh} that the $\gamma$-ray spectrum of 
RX J1713.7-3946, 
corresponds to a flux of CR protons with a energy equal to about 
$0.05\times 10^{51}$ erg (consistent with Ginzburg \& Syrovatskii hypothesis) 
colliding with a molecular cloud of approximately 300 solar masses, 
which represents a quite favorable situation. 
In other words, it difficult to imagine CR fluxes in SNRs much more 
energetic than this or a much larger target mass. 
SNRs can be closer than RX J1713.7-3946 which has a distance from us equal to about 
$D\sim 1 {\rm kpc}$. Keeping the same parameters, this would increase 
the $\gamma$ and $\nu$ fluxes as $D^{-2}$ and, correspondingly, the 
 expected signal in $\nu$ telescopes. However, there will be no gain with respect to the 
atmospheric $\nu$ background, if the linear dimension $L$ of the sources is the same as 
RX J1713.7-3946, since the angular dimension $\theta=L/D$ 
increase with decreasing distances. For this reason, the best candidates are 
close and young SNRs which had no time to expand to large radii 
and which are also expected to produce spectra which 
extend to larger energies, 
having the most energetic proton less time to escape.
A promising candidate could be the 
young and close SNR named Vela Jr (angular size $2^\circ$), 
as suggested by the extrapolations of the first H.E.S.S.\ 
observations \cita{palp,Villante:2007mh,Vissani:2008zz}.

\section{\sf\color{verdon}Events in neutrino telescopes}
\label{rates}

%
\begin{figure}[t]
\par
\begin{center}
\includegraphics[width=8.cm,angle=0]{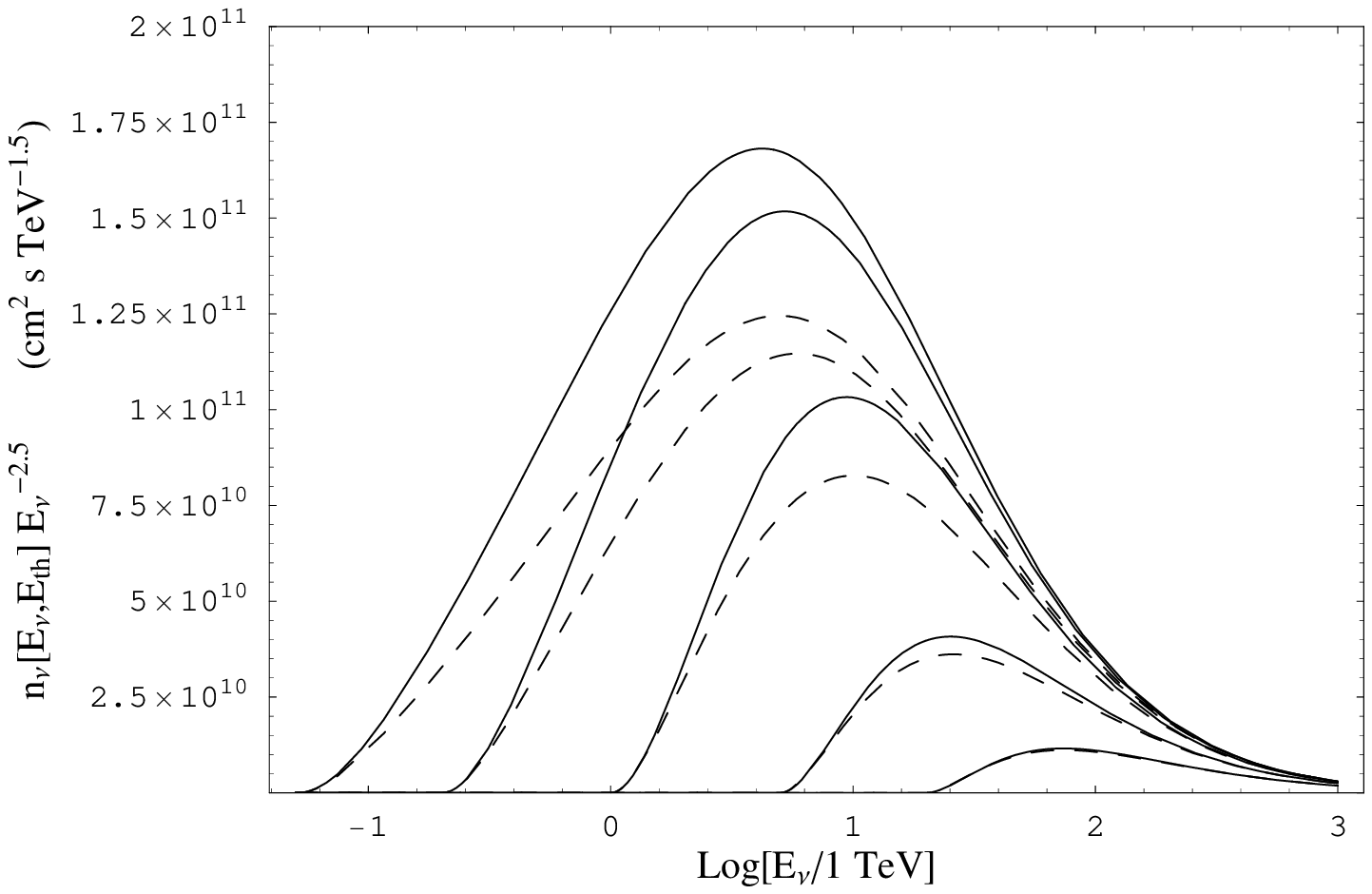}
\includegraphics[width=8.cm,angle=0]{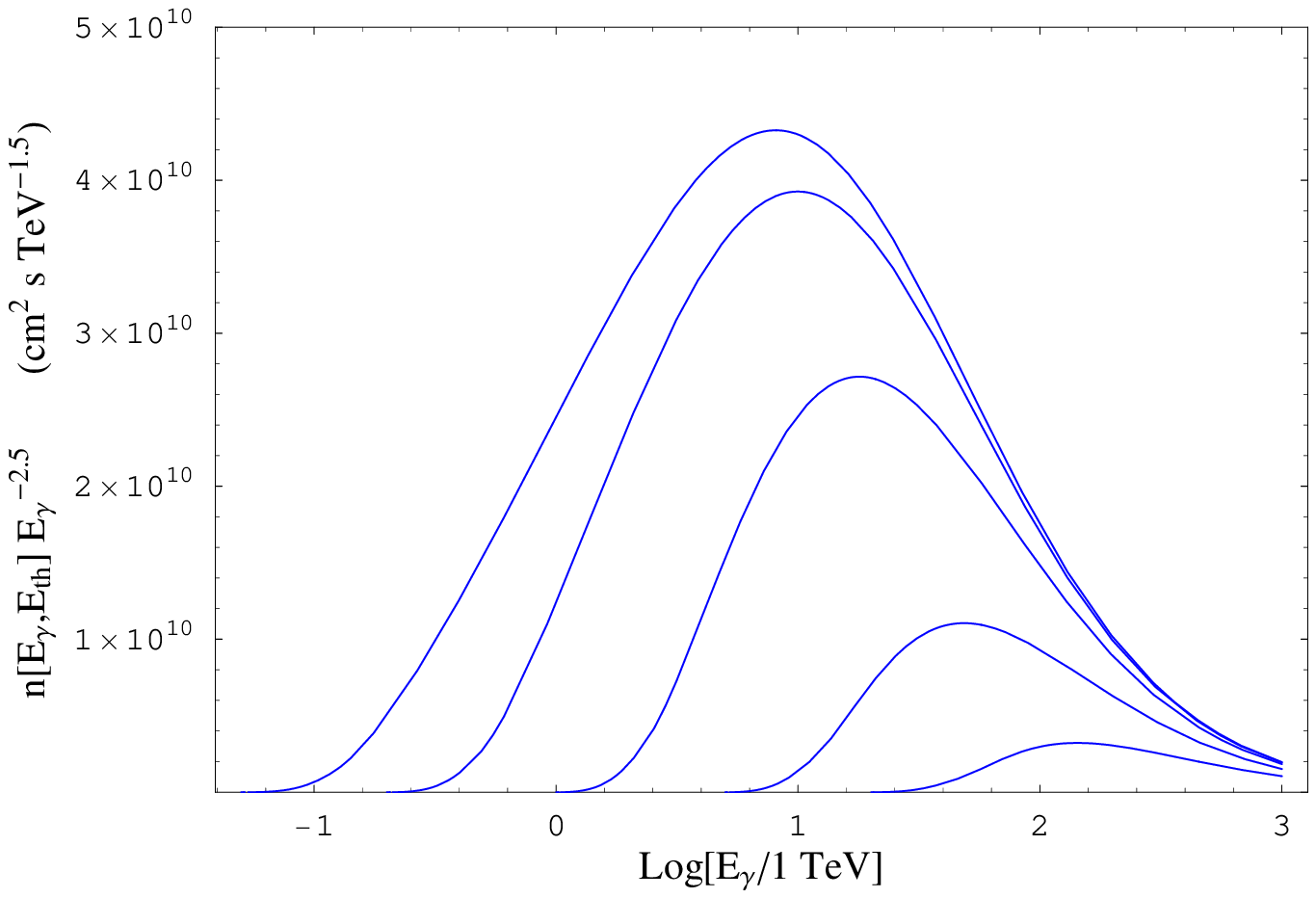}
\end{center}
\par
\vspace{-5mm} \caption{\em {\protect\small {\sc Left Panel:} The function $n_{\nu_{\mu}}[E_\nu,E_{\rm th}] E_\gamma^{-\alpha}$ (solid) and $n_{\overline{\nu}_{\mu}}[E_\nu,E_{\rm th}] E_\gamma^{-\alpha}$ (dotted) for $\alpha=2.5$ and $E_{\rm th}= 0.05,\;0.2,\;1,\;5,\;20$ TeV. {\sc Right Panel:} The function $n[E_\gamma,E_{\rm th}] E_\gamma^{-\alpha}$ for $\alpha=2.5$ and $E_{\rm th}= 0.05,\;0.2,\;1,\;5,\;20$ TeV.\label{gammanuresponse}}}
\end{figure}
%

Charged-current interactions of $\nu_\mu$ and $\overline{\nu}_\mu$ 
produce muons and antimuons that can be observed in neutrino telescopes. 
The number of muons and antimuons reaching an area $A$ 
and in a time of observation $T$ is:
\begin{equation}
N_{\mu+\overline{\mu}} = 
f_{\rm liv} \cdot A \cdot T \cdot 
\int_{E_{\rm th}}^\infty dE \; \Phi_{\nu_\mu}[E]\times
Y_\mu[E,E_{\rm th}]( 1 -\overline{a}_{\nu_\mu}[E] ) + (\nu_\mu \rightarrow \overline{\nu}_\mu)
\end{equation}
where $E$ is the energy of the neutrino at the point of interaction and $E_{\rm th}$ is the
energy threshold for muon detection. Following \cita{Costantini:2004ap}: 
{\em (1)} we assume $A=1 {\rm km}^2$ and $T=1$ solar year;
{\em (2)} we consider that the SNR RX J1713.7-3946 is visible for $f_{\rm liv}= 78 \%$ of the sideral year, as in Antares location;
{\em (3)} the neutrino absorption coefficient $\overline{a}_{\nu_\mu}(E)$, averaged over the 
daily location of the source, is calculated for standard rock; 
{\em (4)} the probability 
$Y_\mu(E,E_{\rm th})$ that a neutrino of energy $E$ produce a muon that reach the 
detector with an energy larger than $E_{\rm th}$ is calculated for water. 

The above equation can be rewritten in the form:
\begin{equation}
N_{\mu+\overline{\mu}} = 
\int_{E_{\rm th}}^\infty \frac{dE_\nu}{E_\nu}\;
n_{\nu_{\mu}}[E_\nu,E_{\rm th}]
\; \Phi_{\nu_\mu}[E_\nu] + (\nu_\mu \rightarrow \overline{\nu}_\mu)
\end{equation}
where the function:
\begin{equation}
n_{\nu_{\mu}}[E_\nu,E_{\rm th}] 
= f_{\rm liv} \cdot A \cdot T \cdot 
Y_\mu[E_\nu,E_{\rm th}]\cdot( 1 -\overline{a}_{\nu_\mu}[E_\nu] )\cdot E_\nu \\
\end{equation}
gives the relative contribution of neutrinos with energy $E_{\nu}$ to the total event rate above 
a detection threshold $E_{\rm th}$. In the left panel of Fig.~\ref{gammanuresponse} we show the function 
$n_{\nu_{\mu}}[E_\nu,E_{\rm th}]E_\nu^{-2.5}$ as a function of $E_\nu$ for the selected values 
$E_{\rm th}= 0.05, \;0.2, \; 1,\; 5,\; 20 $ TeV. 
One sees that the bulk of the signal is expected
to arise from the spectral region $E_\nu\simeq0.3-100$ TeV which is directly constrained by
photon observational data, provided that the detection threshold is below few TeV.
 
The previous point can be much better appreciated by rewriting 
the expected event rate in $\nu$-telescopes in terms of the observed photon flux. 
By using the results of the previous sections, we obtain: 
\begin{equation}
N_{\mu+\overline{\mu}}=\int_{E_{\rm th}}^{\infty}\frac{dE_\gamma}{E_\gamma} \; n[E_\gamma,E_{\rm th}]\Phi_\gamma[E_\gamma]
\label {mufromgamma}
\end{equation}
where the function $n[E_\gamma,E_{\rm th}]$ is given by:
\begin{equation}
n[E_\gamma,E_{\rm th}]= 
\int_{E_{\rm th}}^{E_\gamma} \frac{dE_\nu}{E_\nu} \; K^{\rm osc}_{\nu_\mu}[E_\nu/E_\gamma]\,
n_{\nu_{\mu}}[E_\nu,E_{\rm th}] + (\nu_\mu \rightarrow \overline{\nu}_\mu)
\end{equation}
and $K_{\nu_\mu}^{\rm osc}[x]$ is the photon-neutrino kernel given in Eq.~(\ref{kernel}).
The function $n[E_\gamma,E_{\rm th}] E_\gamma^{-2.5}$ is shown as a function of $E_\gamma$
in the right panel of Fig.~\ref{gammanuresponse}. We see immediately that
the spectral region probed by H.E.S.S. data 
({\em i.e.}, $E_\gamma=0.3-300\ {\rm TeV}$) is the 
most relevant energy region to derive expectations for future neutrino telescopes
when the detection threshold is $E_{\rm th}\le 1$ TeV.
In this case, neutrinos that effectively contributes to the $\nu-$telescopes signals 
are connected to photons which are observed with \v{C}erenkov  $\gamma-$ray telescopes. 
If the threshold is much larger than this, on the contrary, the function 
$n[E_\gamma,E_{\rm th}] E_\gamma^{-2.5}$ is peaked outside the probed region
and one is forced to rely on theoretical extrapolations.

By following a procedure analogous to that adopted in the previous section, we can calculate the expected signal 
from RX J1713.7-3946 directly from photon observational data. If the photon flux is approximated as in Eq.~(\ref{rawdata}), we obtain: 
\begin{equation}
N_{\mu+\overline{\mu}}= \sum_j \varphi_j \, \overline{n}_j[E_{\rm th}]
\label{ratesrawdata}
\end{equation}
where $\varphi_j=\Phi_j\cdot E_j^\alpha$ is the observational value for the (rescaled) photon flux in the $j-$th energy bin and
the factors:
\begin{equation}
\overline{n}_{j}[E_{\rm th}]=\int \frac{dE_\gamma}{E_\gamma}\; 
n[E_\gamma,E_{\rm th}]\,E_\gamma^{-\alpha} \, W_j[E_\gamma]
\end{equation} 
``weights" 
the contribution of each energy  bin to the signal above the threshold $E_{\rm th}$
($W_j[E_\gamma]\equiv1$ for $E_{j,\rm inf}\le E_\gamma \le E_{j,\rm sup}$ and zero elsewhere).
The uncertainty in the predicted neutrino signal can be evaluated by propagating linearly
the observational errors $\Delta\varphi_j$, obtaining:
\begin{equation}
\Delta N_{\mu+\overline{\mu}} = \sqrt{\sum_j  \Delta\varphi_j^2  \,\overline{n}_j[E_{\rm th}]^2}
\end{equation}

\begin{table}
\begin{center}
\caption{\em {\protect\small The number of neutrino events, $N_{\mu+\overline{\mu}}$, from SNR RX J1713.7-3946 expected in Antares location 
per km$^2$ per year for various energy thresholds, according to H.E.S.S $\gamma$-ray data. 
The uncertainty of the predicted signal, $\Delta N_{\mu+\overline{\mu}}$,
is obtained by propagating H.E.S.S. observational errors. The atmospheric neutrino background, $N^{\rm Atmo}_{\mu+\overline{\mu}}$, 
is estimated from \cita{lipariatmo} (see text for details).}} 
\vspace{0.5 cm}
\begin{tabular}{c|cc|c|r}
$E_{\rm th}$	(TeV)		& $N_{\mu+\overline{\mu}}$   & $\Delta N_{\mu+\overline{\mu}}$ & $\frac{\Delta N_{\mu+\overline{\mu}}}{N_{\mu+\overline{\mu}}}$	 	& $N^{\rm Atmo}_{\mu+\overline{\mu}}$		\\
\hline
0.05 & 5.65 & 0.35 & 0.06 & 20.5\\
0.2  & 4.67 & 0.33 & 0.07 & 6.6 \\
1   & 2.44 & 0.28 & 0.11 & 1.1\\
5    & 0.57 & 0.17 & 0.30 &  0.1\\
20     & 0.08 & 0.07 & 0.95 & 0.007\\
     \hline
\end{tabular}
\label{tabella2}
\end{center}
\end{table}

 The number of neutrino events from SNR RX J1713.7-3946
in Antares location after one year of data taking is reported in
Tab.~\ref{tabella2}. 
We see that a km$^2$ class neutrino telescope  will be able to collect few events 
per year, if the threshold will be lower than about 1 TeV. 
The {\em observational} uncertainty is less than 10\%, indicating that  
H.E.S.S. data very well constrains the expected neutrino signal. 
The above estimates do not include possible contributions from the low 
($E_\gamma\le 0.3$ TeV) and high energy ($E_\gamma\ge 300$TeV) tails of the photon flux,
which are not constrained by observations. However, these contributions are expected to
be negligible (at the few per cents level), if the threshold is lower than few TeV.
 We remind that our calculation is affected by $\sim 20\%$ {\em systematic} error
due to uncertaintie in hadronic cross sections, neutrino oscillation 
parameters and the approximations implicit in our method, as conservatively estimated 
in the previous sections.

 The above results show that the detection threshold will be a crucial parameter.
Clearly, it cannot be much larger than 1 TeV, otherwise the event rate would be negligible. 
However, it cannot be much lower than this, due the presence of the atmospheric neutrino background.
At present, the atmospheric neutrino fluxes have a quite large uncertainty due to 
imprecise knowledge of primary CR spectra at earth, of hadronic cross sections 
and the poor understanding of charmed particles production (see \cita{lipariatmo}). 
However, in the next future, they will be measured by neutrino telescopes.
 
In order have feeling of the background levels, 
we approximate the atmospheric neutrino fluxes according to 
$\Phi_{\nu_\mu}^{\rm Atmo}[E_\nu]=\Phi_{\overline{\nu}_\mu}^{\rm Atmo}[E_\nu]=C\cdot (E_\nu/1 {\rm TeV})^{-3.6}$,
as it is appropriate in the energy range of our interest.
The normalization constant $C=1.8\times\,10^{-11} {\rm TeV}^{-1}\,{\rm cm}^{-2}\,{\rm s}^{-1}$ has been 
obtained by fitting the vertical neutrino flux of \cita{lipariatmo} 
(which underestimates the average atmospheric neutrino flux in the direction of the SNR)  
and integrating over an angular window $\theta=1^\circ$. 
By using this parametrization, we obtain the number of events given in the last column of Tab.~\ref{tabella2}. 
Clearly, these values are purely indicative.
They allow us, however, to conclude that there is no real 
gain to lower the threshold 
below the TeV level, in agreement with the finding of 
\cita{liparinim}. In conclusion, 
the best choice seems $E_{\rm th} \sim 1$ TeV, 
since it allows us to probe neutrino emission: 
{\it i)} in an energy region very-well constrained by $\gamma$-ray data; 
{\it ii)} collecting few events per year;
{\it iii)} with a signal-to-background ratio order one or larger.

\section{\sf\color{verdon}Summary}
\label{conclusions}

Future observations in neutrino telescopes have the potential to test an 
important theoretical paradigm, that the young SNRs are the main site of 
acceleration of the galactic cosmic rays. At present, it is useful to use 
the existing observations of very high energy gamma rays from SNRs to derive 
quantitative predictions for neutrinos from these sources. The main 
results obtained in this paper are the following:


{\em i)} We have discussed a conceptually and computationally simple method to 
extract precise predictions for neutrinos from supernova remnants and from 
other hypothetical sources, transparent to their gamma rays. This method 
(that is based and that elaborates on our previous work on the 
subject~\cita{Vissani:2006tf,Costantini:2004ap,Villante:2007mh,Vissani:2008zz}) is superior to other ones present in the 
literature, in that it does not need a preliminary parametrization of the 
gamma ray observations and, as we demonstrated, permits one to use directly 
the gamma ray data as an input.

{\em ii)} Our method allows us to propagate easily the $\gamma$-ray observational errors
and, thus, to understand how well neutrino fluxes and the signal in future neutrino
telescopes can be constrained by $\gamma$-ray observations. Our analysis, and 
specifically the application to the best studied SNR (RX J1713.7-3946) shows that 
the present, successful program of observations of very high energy gamma rays 
from certain supernova remnants (in particular, with imaging arrays of 
\v{C}erenkov telescopes) covers the right energy region to derive 
expectations for the forthcoming neutrino telescopes. As an example, the signal produced in 
$\nu$-telescopes by RX J1713.7-3946 is predicted with an observational uncertainty equal to about $\sim 10\%$.

{\em iii)} We have discussed the sources of theoretical and systematic uncertainties 
in our calculation. In particular, we have checked that the error arising from unknown neutrino 
oscillations parameters is very small (at the level of $\sim2\%$), as a result of the partial cancellation 
of the (much larger) anti-correlated contributions of electron and muon neutrino oscillation probabilities
to the total error budget.\footnote{This is also due the vicinity of a physical boundary, almost saturated by oscillations, as 
can be seen from in Fig.~\ref{Fig1}.}
The main source of uncertainty in the predictions is due to the modeling of 
the hadronic interactions and was 
conservatively estimated at $\sim 20\%$ level. 
This error could be more precisely assessed after a systematic comparison of existing numerical
codes and overview of the available experimental data. 

{\em iv)} It seems possible, at least for the best observed SNR (RX 
J1713.7-3946), to succeed and detect a neutrino signal with exposures of 
the order of year$\times$ km$^2$, provided that the detection threshold in future
neutrino telescopes will be equal to about $\sim 1$TeV. Another promising and perhaps better 
young SNR is Vela Jr, whose higher part of gamma ray spectrum (above 20 
TeV) is still to be studied.
Due to the presence of the atmospheric neutrino background, it does 
not seem really much useful to lower the threshold for neutrino 
observation below the TeV region. This is in agreement with the
finding 
of~\cita{liparinim}.



In summary, we have shown that it is possible to obtain precise and reliable predictions 
of the neutrino signals expected in neutrino telescopes from SNRs. The observation of 
neutrinos from these objects would amount to a proof of the existence of 
the expected over-density of CR and, thus, to a confirmation of the hypothesis that young
SNRs are the main site of acceleration of galactic CR. 

\section*{\sf\color{verdon}  Acknowledgment}
This work was partially supported 
by the  High Energy Astrophysics 
Studies contract number ASI-INAF I/088/06/0, 
by the MIUR grant for the Projects of National
Interest PRIN 2006 ``Astroparticle Physics'' and by 
European FP6 Network ``UniverseNet'' MRTN-CT-2006-035863.
We thank F.~Aharonian, P.~Blasi, M.L.~Costantini, P.L.~Ghia, 
P.~Lipari, F. Lucarelli and G.~Riccobene for useful discussions.

\newpage



\section*{\sf\color{verdon}  References}
\def\refname{

\end{document}